\documentstyle[12pt]{article}
\begin{document}

\def\beq{\begin{equation}}
\def\eeq{\end{equation}}
\def\bce{\begin{center}}
\def\ece{\end{center}}
\def\bea{\begin{eqnarray}}
\def\eea{\end{eqnarray}}
\def\ben{\begin{enumerate}}
\def\een{\end{enumerate}}
\def\ul{\underline}
\def\ni{\noindent}
\def\nn{\nonumber}
\def\bs{\bigskip}
\def\ms{\medskip}
\def\wt{\widetilde}
\def\wh{\widehat}
\def\Tr{\, \mbox{Tr} \, }
\def\brr{\begin{array}}
\def\err{\end{array}}



\vspace*{3mm}

\begin{center}

{\LARGE \bf
Conformal factor dynamics in the $1/N$ expansion}

\vspace{6mm}

 {\bf E. Elizalde} \\ Department of Physics, Faculty of Science,
 Hiroshima University \\ Higashi-Hiroshima 724,
Japan\footnote{Address june-september 1994; e-mail:
elizalde@fusion.sci.hiroshima-u.ac.jp. On leave of absence
from: Blanes Center for Advanced Studies, CSIC,
and Department ECM and IFAE, Faculty of Physics,
University of  Barcelona, Diagonal 647, 08028 Barcelona,
Spain; e-mail: eli@zeta.ecm.ub.es}\\ and  \\
{\bf S.D. Odintsov} \\
Department ECM, Faculty of Physics,
University of  Barcelona \\  Diagonal 647, 08028 Barcelona,
Spain\footnote{E-mail: odintsov@ebubecm1.bitnet} \\
and Tomsk Pedagogical Institute, 634041 Tomsk, Russia \\

\vspace{15mm}

{\bf Abstract}

\end{center}

We suggest to consider conformal factor dynamics as applying to
 composite
boundstates, in frames of the $1/N$ expansion. In this way, a new model
of effective theory for quantum gravity is obtained. The renormalization
group (RG) analysis of this model provides a framework to solve the
cosmological constant problem, since the value of this constant
 turns out to be suppressed,
as a result of the RG contributions. The effective potential for
the conformal factor is found too.
\vspace{4mm}

\newpage

The investigation of the dynamics of the conformal factor is becoming
very fashionable. At the classical level, conformal-factor
dynamics describe the conformally-flat solutions of the
equations for the gravitation theories. At the quantum level, the
dynamics of the conformal factor (induced by the conformal
anomaly) was suggested in the first paper of Ref. \cite{1} as
a tool for the description of quantum gravity (QG) in the
infrared (IR) phase. Further study of such effective theory for
QG, of its properties, and of some extensions of it, has been
carried out in Refs. \cite{1}.

Conformal factor dynamics give rise to the effective potential
for the conformal factor \cite{2,3}, which is very useful in
QG (for a general review of perturbative QG see \cite{4}). In
particular, for the case R$^2$-gravity (a multiplicatively
renormalizable theory) the corresponding theory for the
conformal factor has been developed in Ref. \cite{5}.

The conformal-factor theory leads naturally to the appearence of
a theory of the four-dimensional sigma model type, with a
very interesting one-loop renormalization \cite{6}. When
interacting with the standard model (SM), the conformal factor
appears as the dilaton of the theory \cite{7}. It may naturally
emerge also in the study of the SM in the context of
non-commutative geometry \cite{8}.

There exist a certain number of quantum field theories (for
example the four-fermion models \cite{9,10}, for a recent
discussion see \cite{11}) which allow for an analytical study of
their composite boundstates. Some aspects of the gravitational
interaction with the Nambu-Jona-Lasinio (NJL) model have been
studied in Refs. \cite{12}-\cite{14} already. It is the purpose of the
present letter to investigate specifically conformal factor dynamics
as corresponding to composite boundstates, using the $1/N$ expansion
and drawing some analogies from the four-fermion theories
 \cite{9}-\cite{10}.

Our starting point is the two-dimensional theory with action
\beq
S = \int d^2x \, \sqrt{-g} \left[ \overline{\psi} \left( i
\gamma^\mu (x) \nabla_\mu -m \right) \psi + R - \frac{
\Lambda}{2} \right], \label{1}
\eeq
where the massive $N$-component spinor $\psi$ is considered to be
a quantum field. The gravitational field,  on the other hand,
may be either classical or quantum. We also consider the
conformal parametrization of the metric
\beq
g_{\mu\nu} = \rho^2 \eta_{\mu\nu}, \label{2}
\eeq
where $\rho$ is the conformal factor (in general it is $\rho =
\rho (x)$) and $\eta_{\mu\nu}$ is  the flat fiducial metric.
In the QG case, the choice (\ref{2}) corresponds to the gauge
fixing. Substituting (\ref{2}) into (\ref{1}) one gets, at the
clasical level,
\beq
S = \int d^2x \left[ \overline{\chi} \left( i \gamma^\mu \partial_\mu
- m \rho \right) \chi - \frac{\Lambda}{2} \, \rho^2 \right],
\label{3}
\eeq
where $\chi = \rho^{1/2} \psi$. Rescaling $\rho \rightarrow
\Lambda^{1/2} \rho$, we get
\beq
S= \int d^2x \, \left[ \overline{\chi} \left( i \gamma^\mu \partial_\mu
- h \rho \right) \chi - \frac{1}{2} \rho^2 \right],  \label{4}
\eeq
where $h= m \Lambda^{-1/2}$. As one can see, action (\ref{4}) has the
form that is typical for the Gross-Neveu (GN)
 model ($\rho \sim \overline{\chi}
\chi$). The dynamics of this model are quite well known \cite{10}:
asymptotic freedom in the UF limit
\beq
h^2(t) = \frac{h^2}{1 + h^2 N t /\pi},  \label{5}
\eeq
where $t$ is the RG parameter.
However, since (\ref{4}) describes also the dynamics of the conformal
factor, the interpretation of the function $h^2(t)$ is now completely
different from the original interpretation. The $h^2(t)$ here
 is a combination of  the fermionic mass in (\ref{1})
 and of the two-dimensional cosmological
constant $\Lambda$. Using the anomalous scaling dimension one gets
the running composite field
\beq
\rho (x,t) = \rho (x) \left( 1 + h^2 N\, t/\pi \right)^{-1/2}.
\label{6}
\eeq
The conformal factor has acquired the classical dimension after
the rescaling $\rho \rightarrow \Lambda^{1/2} \rho$. Hence, one may
argue that the $t$ dependence is due completely to that of the
cosmological (dimensional) constant, i.e., $\Lambda (t) \sim
\Lambda \ (1 + h^2 Nt /\pi )^{-1}$. Therefore, there appears to
be a screening of the cosmological constant in the $1/N$
expansion in the UF regime.

One can also investigate specific features of
 the effective potential for the conformal factor,
which coincides with the GN effective potential \cite{10}.
In particular, the appearence of a minimum, i.e., a  non-zero
vacuum expectation value (v.e.v.),  for the conformal factor
\beq
\rho = \rho_0 \exp \left( 1 - \frac{\pi}{h^2N} \right)  \label{7}
\eeq
is interesting. After having shown the possibility to study the dynamics
of the conformal factor in two dimensions as the dynamics of the GN
model, we now turn to the four-dimensional theory, which is
physically
more interesting.

We start from the multiplicatively renormalizable theory \cite{4}
with  action
\beq
S = \int d^4x \, \sqrt{-g} \, \left[ \overline{\psi}
\left( i \gamma^\mu (x) \nabla_\mu - m \right) \psi - \frac{
\Lambda}{\kappa^2} -\frac{R}{\kappa^2} + \frac{W}{\lambda_1}
-\frac{UR^2}{3\lambda_1} \right], \label{8}
\eeq
where $\psi$ is an $N$-component spinor and $W$ the square
of the Weyl tensor. The gravitational field may be choosen to be
classical or quantum, and the theory remains multiplicatively
renormalizable in both cases.

We  shall work again with the conformal parametrization (\ref{2})
for the four-dimensional metric. In the case of four-dimensional
QG this does not fix the gauge, contrary to what happens in two
dimensions, but it can
still be considered as a convenient background \cite{5}.
Rewriting action (\ref{8}), we get
\bea
S&=& \int d^4x \, \left\{ \overline{\chi} \left( i \gamma^\mu
\partial_\mu - m\rho \right) \chi -\frac{\Lambda}{\kappa^2}
\rho^4 - \frac{6}{\kappa^2} (\partial \rho )^2 \right. \nn \\
&& \hspace{2cm} \left. +
\frac{12U}{\lambda_1} \left[ \sigma \Box \sigma + 2 (\partial
\sigma )^2 \Box \sigma + (\partial \sigma)^2 (\partial
\sigma)^2 \right] \right\},  \label{9}
\eea
where $\chi =\rho^{3/2}\psi$ and $\sigma = \ln \rho$.
In this way we have got the classical theory for the conformal
factor. At the quantum level, the theory (\ref{9}) may be
considered as an effective theory for QG (see also
\cite{1}-\cite{3} and \cite{5}).
If we drop the $\rho$-terms with derivatives from action
(\ref{9}), we obtain a model that is reminiscent of the
NJL model (where, of course, owing to the absence of the
$M^2\rho^2$-term, it is $\rho \sim (\overline{\chi} \chi
)^{1/3}$).

Now we are going to study the theory (\ref{9}) in the large-$N$
limit, while concentrating our attention
on the RG and low-derivative terms
in (\ref{9}). The higher-derivative terms are actually of
lesser importance, moreover, they simply disappear in the
subsequent analysis of the effective potential for the conformal
factor.
First of all, we rescale $\rho \rightarrow \sqrt{12} \,
\rho /\kappa$ and denote $h=m\kappa /\sqrt{12}$ and $\lambda =
\Lambda \kappa^2 /6$. Thus,
\beq
S= \int d^4x \, \left[ \overline{\chi} \left( i\gamma^\mu
\partial_\mu - h\rho \right) \chi -\frac{\lambda}{24}
\rho^4 - \frac{1}{2} (\partial \rho )^2 \right].
\label{10}
\eeq
By integrating over the fermionic field, we get the effective potential
for the conformal factor at large $N$:
\beq
V(\rho ) = \frac{\lambda \rho^4}{24} + iN \Tr \ln \left( i \gamma^\mu
\partial_\mu - h\rho \right),  \label{11}
\eeq
where $\rho$ is  constant. Supposing that, as usually, for large $N$
$\lambda$ scales as $\lambda \sim \wt{\lambda} N$, where $\wt{\lambda
}$ does not depend on $N$, and using a finite cut-off $\mu$ (see
also \cite{13}) we get (notice that $N$ has been factored out)
\beq
V(\rho )= \frac{\wt{\lambda} \rho^4}{24} -\frac{1}{(4\pi)^2} \left[
\rho^2 \mu^2 + \mu^4 \ln \left( 1 + \frac{\rho^2}{\mu^2} \right) -
\rho^4 \ln \left( 1 + \frac{\mu^2}{\rho^2} \right)\right].
\label{12}
\eeq

The v.e.v. of the conformal factor can be found from (\ref{12}) as
the solution of the equation
\beq
\frac{\partial V(\rho)}{\partial \rho} = \frac{\wt{\lambda}}{6}
\, \rho^2
- \frac{1}{(2\pi)^2} \left[ \mu^2 -\rho^2 \ln \left( 1+
\frac{\mu^2}{\rho^2} \right) \right] = 0.
\label{13}
\eeq
In Table 1 we present a sample of numerical values of the solution
$\rho^2 / \mu^2$ (which lie between 0 and 1 for $0.04644 \leq
\wt{\lambda} \leq 1$).
\begin{table}

\begin{center}

\begin{tabular}{|c|c|}
\hline \hline
 $\wt{\lambda} $  & root $\rho^2 / \mu^2$ \\
 \hline \hline $ 10^{-4}$ & $27.2361$ \\
\hline  $10^{-3}$ & $8.3933$ \\
\hline  $0.01$ & $2.4517$ \\
\hline  $0.464$ & $1.00043$ \\
\hline  $0.465$ & $0.99912$ \\
\hline  $0.05$ & $0.9558$ \\
\hline  $0.1$ & $0.6166$ \\
\hline  $0.2$ & $0.3853$ \\
\hline  $0.3$ & $0.2880$ \\
\hline  $0.4$ & $0.2326$ \\
\hline  $0.5$ & $0.1962$ \\
\hline  $0.6$ & $0.1702$ \\
\hline  $0.7$ & $0.1506$ \\
\hline  $0.8$ & $0.1353$ \\
\hline  $0.9$ & $0.1229$ \\
\hline  $0.99$ & $0.1137$ \\
\hline  $1$ & $0.11275$ \\
\hline \hline \end{tabular}
\caption{ {\protect\small Numerical values for the vacuum expectation
value of the conformal factor, corresponding to a sample of values
of  $\wt{\lambda} $ between 0 and 1. They are obtained as the root
of Eq. (\ref{13}) that lies between 0 and 1 (notice that
for  $ \wt{\lambda} \leq 0.04643 $ there is no root in that interval). } }

\end{center}

\end{table}
Notice that from the point of view of the original theory, a non-zero
v.e.v. for $\rho$ is more acceptable physically, because for $\rho =0$
the conformal parametrization  (\ref{2}) becomes degenerate.

Now we turn to the study of the renormalization structure of theory
(\ref{10}),
which is rather non-trivial. There are a few different ways to renormalize this
theory, the actual problem being the fact that $\rho$ is dimensionless.
Considering (\ref{10}) in the large-$N$ approximation, one can perform the
change $\rho \rightarrow i\rho$, which introduces an overall minus sign
in the parametrization (\ref{10}), producing a kinetic term of the standard
form. The factor $i$ in the fermionic sector can be absorbed in $h$
(imaginary fermion mass). Then, supposing that $\rho$ is also a quantum
field  (i.e. that gravity is quantized in (\ref{18})),
the renormalization of such theory, of Yukawa type, can be done
in the standard way. The beta functions
are calculated to be
\beq
\beta_h = \frac{4Nh^4}{(4\pi)^2}, \ \ \ \ \beta_\lambda =
\frac{3\lambda^2 +8N\lambda h^2 - 48 N h^4}{(4\pi)^2}.  \label{14}
\eeq
We see that $h=\lambda =0$ is the IR stable fixed point, and in the IR
regime ($t \rightarrow - \infty$), we have
\beq
h^2(t) \sim -\frac{4\pi^2}{Nt}, \ \ \ \ \ \ \lambda (t) \sim - \frac{
48 \pi^2}{Nt}.  \label{15}
\eeq
Supposing that the Newton coupling constant is a non-running coupling, then
the running of $h^2(t)$ gives the RG screening of the imaginary fermionic mass
in the infrared. At the same time, the cosmological constant quickly decays
in the IR. Thus, in the considered phase of the effective theory for the
conformal factor, we get a solution of the cosmological constant
 problem, as a result of RG effects in the IR.

We may also consider a different version of (\ref{10})
---as it stands--- which will correspond eventually to another
QG phase. Now there are no imaginary
parameters. After renormalization of $\rho$ (taking into account the
negative sign for $(\partial \rho )^2$), we obtain
\beq
\beta_h = -\frac{4Nh^4}{(4\pi)^2}, \ \ \ \  \beta_\lambda =
\frac{3\lambda^2 -8N\lambda h^2 - 48 N h^4}{(4\pi)^2}.  \label{16}
\eeq
The theory has now an UF stable fixed point ($t \rightarrow +\infty$),
 where the
behavior of $h^2$ and $\lambda$ is
\beq
h^2(t) \sim \frac{4\pi^2}{Nt}, \ \ \ \ \ \  \lambda (t) \sim -\frac{48
\pi^2}{Nt}.   \label{17}
\eeq
Hence, we now obtain a decrease of the cosmological constant in the UF limit.
Notice that in comparing this theory with the standard scalar self-interacting
theory, we do not have here physical restrictions on the sign of $\lambda$,
and a negative sign is perfectly acceptable.

Finally, let us observe that from the point of view of Eq. (\ref{9}), where
$\rho$ is dimensionless, one can understand (\ref{9}) as being a kind of
four-dimensional $\sigma$ model \cite{6}. Then, it has sense to discuss
a renormalization of (\ref{9}) of $\sigma$-model type, at large $N$. In
that case $\lambda_2 = \Lambda /(6 \kappa^2)$, $\rho$ (rescaled as $\rho
\rightarrow \sqrt{12}\, \rho$) and $m$ are not renormalized to leading
order in $N$, and
\beq
\beta_{\lambda_2} = \frac{3\lambda_2^2 -48 N m^4}{(4\pi)^2}, \ \ \ \
\beta_{\kappa^{-2}} = \frac{Nm^2}{4\pi^2}.  \label{18}
\eeq
As a result one can see that $\kappa^{-2} (t) \sim Nt$ and $\lambda_2 (t)
\sim -Nt$ at large $t$. Hence, considering the gravitational constant as the
running coupling constant and using the generalized RG we do not obtain a
damping of the running cosmological constant in that case.

It is very interesting to observe that if we would have started
 from the anomaly-induced theory for the cosmological factor \cite{1}
 and had considered its interaction with the massive fermion, we would
have obtained the same theory  (\ref{9}) (only some relative
 coefficients of the higher-derivative terms
would be different). Hence, the preceding discussion can be applied
without the least change  to this case too.

As next step we  will now consider the appearence of the conformal factor in
the proper NJL dynamics.   Starting point is theory (\ref{8}),
where in the
place of the fermionic sector we substitute the action for the NJL model
\cite{11,12}
\beq
L= i \overline{\psi} \gamma^\mu (x) D_\mu \psi + (\overline{\psi}_L^a
\psi_{Ra} H + \mbox{h.c.} ) - M_\mu^2 H^\dagger H,  \label{19}
\eeq
being $\psi_{L,R} = \frac{1}{2} (1\mp \gamma_5 )\psi$, $H$ an auxiliary
scalar and $M_\mu$  the scalar field mass of the UF scalar $\mu_{UF}$.
Such theory is a four-fermion one \cite{9}, where the four-fermion
coupling constant is $G=M_\mu^{-2}$. Notice that the above theory is
non-renormalizable (effective), so that there is no need, for instance,
 of keeping higher-derivative terms in (\ref{18}). Working in the conformal
 parametrization (\ref{2}), after rescaling $\chi = \rho^{3/2} \psi$
and $H \rightarrow \rho H$ we get the flat-space NJL model, with the only
difference
with respect to \cite{11,12} that the scalar-field mass becomes
$\rho$-dependent: $\wt{M}_\mu^2 = \rho^2 M_\mu^2$ in (\ref{19}). Performing a
block-spin RG transformation (in the $1/N$ expansion) we can study the RG
here precisely in the same way as it was done in \cite{11,12}, the only
difference being that the induced scalar mass will now be
\beq
M_{\mbox{ind}}^2 = \rho^2 M_\mu^2 - \frac{N}{8\pi^2} (\mu_{UF}^2 -
\mu_{IR}^2),  \label{20}
\eeq
and the dilaton sector is not influenced by renormalization. Two parameters,
$\rho$ and $M_\mu^2$, appear in  (\ref{20}), so even the value of the
conformal factor will be defined by the minimization of the
corresponding effective potential, their combination being used for the
approximate cancellation of the corresponding term $N\mu_{UF}^2/ (8
\pi^2)$. Finally, let us point out that starting from a massive model
of GN type
and after using (\ref{2}), the conformal field dynamics are influenced again
by fermionic effects, and the total model includes scalars, as the NJL
model. The study of this sort of model, where the conformal factor is in the
composite field, can be carried out in analogy with that of the NJL model.

In summary, we have considered in this paper the conformal field dynamics
in a framework where the conformal factor becomes naturally a boundstate
and the $1/N$ expansion can be applied. As a particular result, we have
obtained in this way a possible mechanism for the vanishing at the
RG fixed point of the cosmological constant (assuming that the
Newton constant is a non-running coupling). The interaction of the conformal
factor in the NJL model has been discussed too. Our results show that
the $1/N$ expansion can well be applied for the description of
effective theories of QG. A synthesis of this approach with  the
anomaly-induced theory for the conformal factor may be very helpul for the
resolution of the drawbacks of previous scenarios \cite{1} (like
neglecting the spin-2 gravitational modes), by means of the use of the
large-$N$ expansion.

\bigskip

\noindent{\bf Acknowledgments}.

 SDO would like to thank R. Percacci for interesting discussions and
C.T. Hill for correspondence.
EE is grateful to
T. Muta and  the  rest of the members of the Department of Physics, Hiroshima
University, for interesting discussions and warm hospitality.
SDO would like to acknowledge the  hospitality of the
members of the Department ECM, Barcelona University. This work has been
supported  by the Special Exchange Program (Japan), by DGICYT
(Spain),  by CIRIT (Generalitat de Catalunya),
by the Russian Foundation for Fundamental Research, project
no. 94-02-03234, and by ISF, project RI-1000.

\end{document}